\documentstyle[multicol,aps,epsf,here]{revtex}
\tighten
\newcommand{\PostScript}[7]{
\begin{figure}[H]
\begin{center}
\leavevmode
\epsfysize=#1cm
\vspace{#2cm}
\epsfbox{#3}
\par
\parbox{#5cm}{
\vspace{#4cm}
\caption[figure]{\renewcommand{\baselinestretch}{1} \small \normalsize #6}
\label{#7}}
\end{center}
\end{figure}
}

\begin{document}

\newcommand{\half}{\frac {1}{2} }
\newcommand{\eg}{{\em e.g.} }
\newcommand{\ie}{{\em i.e.} }
\newcommand{\etc} {{\em etc.}}

\newcommand{\noi}{\noindent}
\newcommand{\etal}{{\em et al.\ }}
\newcommand{\cf}{{\em cf. }}

\newcommand{\dd}[2]{{\rmd{#1}\over\rmd{#2}}}
\newcommand{\pdd}[2]{{\partial{#1}\over\partial{#2}}}
\newcommand{\pa}[1]{\partial_{#1}}
\newcommand{\pref}[1]{(\ref{#1})}

\newcommand {\be}[1]{
      \begin{eqnarray} \mbox{$\label{#1}$}}

\newcommand{\ee}{\end{eqnarray}}

\def\rdown{\rho_{\downarrow}}
\def\p{\partial}


\title{Current-spin-density-functional
       study of persistent currents in quantum rings}
\vskip 20 mm
\author{S. Viefers} 
\address{Nordita, Blegdamsvej 17, 2100 Copenhagen, Denmark}
\author{P. Singha Deo}
\address{S.N. Bose National Centre of Basic Sciences,
         Calcutta 700091, India} 
\author{S.M. Reimann, M. Manninen and M. Koskinen}
\address{Dept. of Physics, University of Jyv\"askyl\"a, P.O. Box 35,
     40351 Jyv\"askyl\"a, Finland}
\date{\today}

\maketitle

\begin{abstract} 
We present a numerical study of persistent currents in quantum rings 
using current spin density functional theory (CSDFT). 
This formalism allows for a systematic study of the joint effects
of both spin, interactions and impurities for realistic systems. 
It is illustrated that CSDFT is suitable for describing the
physical effects related to Aharonov-Bohm phases by comparing 
energy spectra of impurity-free rings to existing
exact diagonalization and experimental results. 
Further, we examine the effects of a symmetry-breaking impurity
potential on the density and current characteristics of the system
and propose that narrowing the confining potential
at fixed impurity potential will suppress the persistent current
in a characteristic way. 
\end{abstract}
\begin{multicols}{2}

\narrowtext
\section{Introduction}
\label{s1}
Nanoscopic quantum rings small enough to be in the true
quantum limit can nowadays be realized experimentally 
\cite{garcia1,lorke2,lorke1}.
Among the quantum effects manifested in such systems
in the presence of an external magnetic field, is the Aharonov-Bohm (AB)
effect\cite{aharonov1}, leading to periodic oscillations in the energy spectrum
and thus persistent currents. 
This phenomenon, first predicted by Hund\cite{hund1}, was discussed
in connection with superconducting rings\cite{byers1,bloch1} and
more recently predicted to occur also in one-dimensional
metallic rings \cite{buttiker1}. In the ideal case of one electron
in a clean, one-dimensional ring, the Aharonov-Bohm phase picked up by
the electron modifies the periodic boundary conditions, leading to
single particle energies given by
\be{en}
E_n = \frac{\hbar ^2}{2m} \left( \frac{2\pi}{L} \right)^2
                   \left( n - \alpha \right)^2
\ee
where $\alpha = \phi/\phi_0$ is the number of flux quanta penetrating the
ring and $L$ is the length of the ring. The single-particle spectrum
is periodic in $\phi$ with periodicity $\phi_0$.
The persistent current associated with state $n$ is
\cite{byers1,bloch1,buttiker1}
\be{pc}
J_n = -c \frac{\p E_n}{\p \phi}.
\ee
However, in realistic systems, interactions, lateral dimension,
impurities and spin effects
complicate the picture. 
In particular, interactions 
may shift different energy levels relative to each other,
leading to complicated
ground state patterns with transitions between states with different
spin and/or angular momentum as the Aharonov-Bohm flux is increased.
In this way, interactions may decrease the period of the oscillations
in the ground state energy (``fractional Aharonov-Bohm effect'').
The first systematic study of persistent currents in
ideal, one-dimensional metallic rings, including
temperature- and impurity effects but neglecting interactions,
was reported in \cite{cheung1}. Subsequent approaches include
Hubbard model calculations \cite{yu1}, use of Hartree- and 
Hartree-Fock methods \cite{amb1,eckern1}, exact diagonalization
studies \cite{chakraborty1,chakraborty2,niemela1} and very
recently density-functional calculations\cite{emperador1,emperador2}.
Experiments in the early nineties reported observations of
persistent currents in an ensemble of $\sim 10^7$ Cu rings\cite{levy1},
in single gold rings\cite{chandrasekar1} and in a single loop
in a GaAs heterojunction\cite{mailly1}, all in the mesoscopic
range.
Very recently, Lorke \etal\cite{lorke1} reported the first spectroscopic 
data on nanoscopic, self-assembled InGaAs quantum rings containing
only one or two electrons. 

Most of the theoretical approaches mentioned above, have the
limitation that, \eg, interactions, spin effects, impurities
or lateral dimension
had to be neglected to simplify the calculations.  
In this paper, we apply the so-called 
current spin density functional theory (CSDFT) \cite{vignale:rasolt}
including gauge fields in the energy functional.
CSDFT was earlier applied to describe the electronic structure of quantum dots 
\cite{ferconi1,heinonen,ferconi2,steffens,pi,koskinen,reimann,kolehmainen}. 
This method, while being a mean field approach and thus not exact, 
has the advantage that one can take into account all the above effects, 
being more accurate than Hartree-Fock, and it should also be possible 
to take it to higher particle numbers than the exact diagonalization methods 
reported in the literature.
Our aim is, first of all, to examine to what extent the CSDFT
formalism captures the physics due to the Aharonov-Bohm effect,
namely the periodic variations in the energy spectra as function
of flux, and the corresponding persistent currents. Thus,
after introducing the basics of our model in section
\ref{s2}, we present, in section \ref{s3}, the spectra of impurity-free
two- and four particle rings and discuss how they compare qualitatively and
quantitatively to corresponding exact diagonalization- and experimental
results. 
In section \ref{s4} we study the effects of a symmetry breaking
impurity potential on the density profile and persistent current
of six-electron rings. It is also predicted that, for a fixed impurity
strength, narrowing the confining potential will tend to localize the
electrons and suppress the persistent current. Finally, section
\ref{s5} is devoted to discussion and conclusions.

\section{Model and numerical method}
\label{s2}
We consider $N$ electrons of effective mass $m^*$, 
confined to a ring with radius $R_0$ 
by a potential
\be{cpot}
V(r) = \half m^*\omega_0^2 \left( r-R_0 \right)^2.
\ee
When we examine the system in the presence of an impurity, we will introduce
an additional Gaussian potential centered at the bottom of the
potential well $(x=R_0, \ y=0)$,
\be{imppot}
V_I ({\bf r}) 
      = V_0 \ \exp\left( -\frac{(x-R_0)^2}{a^2} - \frac{y^2}{b^2} \right) .
\ee
The AB flux is provided by a flux tube of constant field $B_0$ and 
with radius $r_i$  in the center of the ring; the corresponding
vector potential is chosen as
\be{vpot}
A_{\varphi} &=& \left\{ \begin{array}{c}
        {B_0 r/2, \ \ \ r \leq r_i} \\ \\ 
        {B_0 r_i^2/(2r) , \ \ \ r > r_i}
\end{array} \right. \\
A_r &=& 0 \ \ \ \ ,
\ee
with $r_i$ chosen small enough that the electrons themselves move 
essentially in a field-free region.

Instead of using the quantities $N,~\omega_0$ and $R_0$ to describe the
properties of the system, it is convenient to introduce the average
$1D$ interparticle spacing $r_{s,1D}$ (related to the average density
$\rho_{1D}$ by $r_{s,1D}=1/(2\rho_{1D})=\pi R_0/N$), and the
``degree of one-dimensionality'' $C_F$ \cite{reimann1}. $C_F$ is a
dimensionless number defined as the ratio  between the ``transverse'' 
(oscillator) gap $\hbar\omega_0$ and the Fermi energy; 
the latter is approximated by the Fermi energy of a free, one dimensional
Fermi gas with the same density, 
$\epsilon_F = (\pi^2 \hbar^2 N^2)/(8m^* L^2)$, where
$L$ is the length of the ring. Thus,
\be{cF}
C_F = \omega_0 \ \frac{32 m^* r_{s,1D}^2}{\pi^2\hbar} \ \ \ \ .
\ee 
The higher the value of $C_F$, the narrower is the ring.
We will be using values of $C_F$ for which the system is
essentially described by a single channel,
but with a smooth charge distribution in
space.

For a given set of parameters, we compute the ground state charge- and
current densities using CSDFT. 
In this formalism, originally introduced by Vignale and 
Rasolt \cite{vignale:rasolt}, 
one solves the self-consistent Kohn-Sham type equations, 
\be{KS}
\left[{{\bf p}^2\over 2m^*} + {e\over 2m^*} \left( 
{\bf p} \cdot {\boldmath \cal A}~+~ {\boldmath \cal A}\cdot
{\bf p} \right)
+ {\cal V}_{\delta }\right ]\Psi _{i\delta}=\varepsilon _{i\delta }
\Psi _{i\delta}
\ee
(We have dropped the arguments $\bf r$ for simplicity).
The index $i$ labels the eigenstates with spin 
$\delta = (\uparrow, \downarrow )$,
and 
${\boldmath \cal A}:= {\bf A}+ {\bf A}_{\rm xc}$
and ${\cal V}_{\delta }:= {(e^2/ 2m^*)}A^2 + V_{\delta}
+ V_H+ V_{{\rm xc}\delta }$
are the effective vector and scalar potentials.
Here, $V_H$ is the ordinary Hartree potential and $V_{\delta }
= V +(-) g^*\mu _B B/2$
is the external potential, including the Zeeman energy (which in 
our case is set to zero, as the electrons do not experience the
magnetic field;
$\mu _B=e\hbar /(2m_e)$ is the Bohr magneton).
The exchange-correlation vector and scalar potentials are 
\be{Axc}
e{\bf A}_{\rm xc} = {1\over \rho}
\left\{{\partial\over \partial y} {\partial [\rho e_{\rm xc}(\rho _{\delta },\gamma )]
\over \partial {\gamma }}~ ,~ 
- {\partial\over \partial x} {\partial [\rho e_{\rm xc}(\rho _{\delta },{\gamma })]
\over \partial {\gamma } }\right\}
\ee
and
\be{Vxc}
V_{{\rm xc}\delta }= {\partial [\rho e_{\rm xc}(\rho _{\delta },{\gamma })]\over 
\partial 
\rho _{\delta }}-{e\over \rho } {\bf j}_p\cdot{\bf A}_{\rm xc}~,  
\ee
where 
$\rho $ is the particle density $\rho = \rho_{\uparrow }+\rho_{\downarrow }$ 
with
$\rho _{\delta } = \sum _i | \Psi _{i\delta } |^2$~. 
The paramagnetic current density is given by
$
{\bf j}_p =  -{{i\hbar / (2m^*)}}\sum _{i\delta}
[\Psi_{i\delta}^*\nabla \Psi_{i\delta}
-\Psi_{i\delta}\nabla \Psi_{i\delta}^*]
$,
and the real current density equals
${\bf j}= {\bf j}_p+ {(e / m^*)}{\bf A} \rho $.
The exchange-correlation energy $e_{\rm xc}$ depends on the 
so-called vorticity 
${\gamma } = \left. \nabla \times ({{\bf j}_p / \rho })
\right | _z ~$ of the wave function. 
For the details of the formalism, we refer to Ref.\cite{vignale:rasolt}.
The practical computational techniques that we found necessary to obtain 
convergent solutions of the CSDFT mean field equations, are given in
Ref.\cite{koskinen}.
The parameters and results will be given in effective 
atomic units ($a.u.^*$)
with energy measured in
$Ha^* = 2 Ry^* = m^*e^4/(4\pi\hbar\epsilon\epsilon_0)^2$
and
length measured in
$a_B^* = \hbar^2(4\pi\epsilon\epsilon_0)/m^*e^2$,
where $\epsilon$ is the dielectric constant and $m^*$ the effective
mass. The results can then be scaled to the actual values for
typical semiconductor materials.

\section{Comparison to exact results}
\label{s3}
In order to check whether CSDFT provides a good description of the 
physics related to Aharonov-Bohm oscillations,
we first apply it to some cases where exact diagonalization results are
available, namely impurity-free rings containing two and four electrons,
respectively. 
Exact diagonalization studies of these systems close to the
ideal, one dimensional limit, were presented by Niemel\"a 
\etal\cite{niemela1}, who calculated the energy spectra as function of
the Aharonov-Bohm flux in the presence of interactions. Starting from
the flux-free, non-interacting case, one can roughly understand the 
structure of the full (interacting) spectra from the following effects:
As we have seen, the presence of an AB flux induces persistent currents
and thus favors increasing total  
angular momentum $L$ (with the corresponding sign) of the electrons.
{\em E.g.}, for an even number of particles,
as $\phi$ is increased from zero, $L<0$ states come down in energy
whereas $L\geq 0$ are pushed up in energy. The ground state of the 
non-interacting $N$-electron ring contains a series of crossings 
between different angular momentum states as the flux increases. 
As interactions are turned on, states with the highest possible symmetry
in the spin part of the wave function are favored due to the gain in
exchange energy. Thus, for example, triplet states come down in energy as 
compared to singlet states. This may lead to additional crossovers,
not present in the non-interacting case, 
between states with different spin, in the ground state spectrum.
Hence, as a first step, we proceed to check to what extent CSDFT
can produce these features. \\
We start by considering a ring with two electrons, choosing a 
radius $R_0 = 1.5$ and $\omega_0=1.0$. This corresponds to the actual
values estimated for the experimentally realized rings recently reported
by Lorke \etal\cite{lorke1}. Figure \ref{f1} shows the ground state energy
per particle
of the two-electron system vs. $\phi/\phi_0$ as computed from CSDFT.
(Note that, for this choice of parameters, the electron density does
not go entirely to zero in the center of the ring, so the electrons get
partly exposed to the external field. This causes the energy
spectrum to tilt upwards instead of being strictly periodic in $\phi$,
in contrast to, \eg, Fig.\ref{f3}, where the ring radius is larger.) \\
For the lowest values of $\phi$, the $S=0$, $L=0$ state is lowest in
energy, with $S$ denoting the spin. 
Then a crossover takes place to the triplet state with $L=-1$
and then back to $S=0$ with $L=-2$. Exactly the same features were
found in the exact diagonalization study \cite{niemela1}, though for a
more one-dimensional ring. The main difference 
between the two methods is that the cusp 
corresponding to the transition between $L=0$ and $L=-2$ in the $S=0$
state (which is not the ground state here) near $\phi/\phi_0=1/2$ is
rounded off in the mean field calculation, and the transition between
the different $L$ states is gradual. The reason is the explicit breaking
of the rotational symmetry in the internal structure of the wave function
that is mapped out by the self-consistent mean-field solution. 
Due to this symmetry breaking, the angular momentum is no longer
a ``good'' quantum number, and non-integer $L$-values
are allowed. 
On the other hand, cusps at transitions between
different {\em spin} states are not smeared out by the LSDA. \\
Furthermore, one can also check by direct comparison with the experimental 
results of Lorke \etal\cite{lorke1}
that even quantitatively, our approach
provides reasonable predictions: 
The energy per particle in the 
two-electron ring at zero external flux can be roughly estimated from the 
experimental data to be around $14$meV.
Converting our calculated result from effective to physical units,
using the estimated values $m^* = 0.07 m$ and $\epsilon = 12.6$, the
energy per particle at zero flux is about $9$meV. \\
The good agreement between CSDFT and the exact results is encouraging
and perhaps not too surprising:
It has previously been shown by Ferconi and Vignale \cite{ferconi1} 
that CDFT provides 
an astonishingly accurate
description of even very small quantum dots, and the same seems 
to be the case here, in the presence of spin. Similar conclusions 
were reached in a recent, independent, density-functional study 
by Emperador \etal\cite{emperador2} of the same two-electron rings.
\PostScript{6}{0}{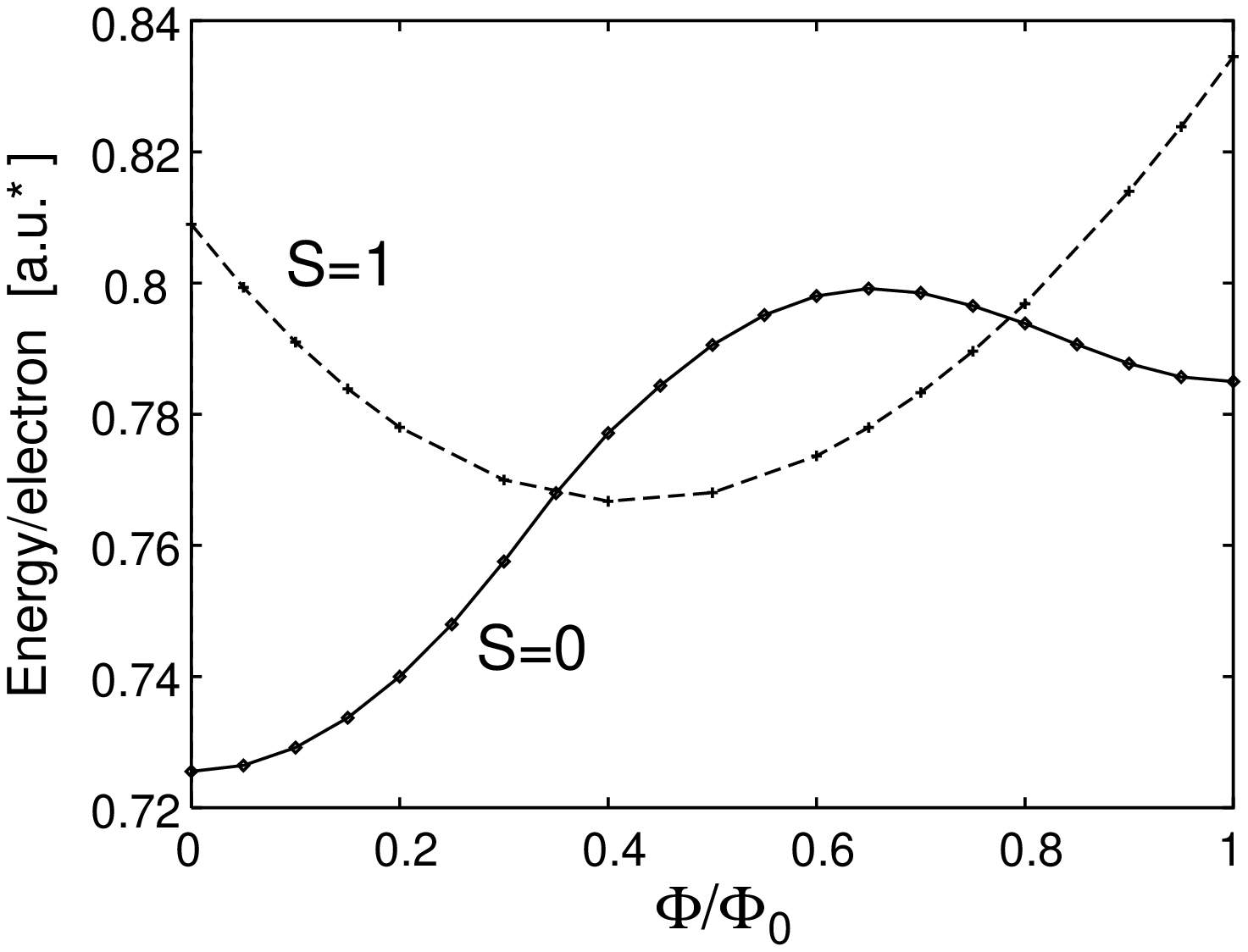}{0.5}{14}{
Two-particle spectrum: Energy per electron vs. flux of the
$S=0$ (solid line) and $S=1$ (dashed line) states for a
two-electron ring with $r_{s,1D}=2.4$, $C_F=18$, corresponding
to the experimental parameters in \cite{lorke1}.}{f1} 
Next, we consider a four-electron ring with $r_{s, 1D}=2.5$, $C_F = 10$. 
Fig.\ref{f2} shows the energy per particle
for the lowest $S=0$ and $S=1$ states, respectively, as computed from CSDFT.
We see that the singlet state is always the ground state, except for 
$\phi/\phi_0$ near $0$ and $1$, where the two states are nearly degenerate.
For both $S=0$ and $S=1$, the total angular momentum changes
continuously from $L=0$ at zero flux via $L=-2$ at $\phi / \phi_0 = 1/2$ to
$L=-4$ when one flux quantum penetrates the ring. These features agree well
with the exact diagonalization results in \cite{niemela1}, which again were
performed for a more one-dimensional ring.  \\
We note in passing that changing the 
flux, keeping all other parameters fixed, can introduce a
weak spin- or charge density wave in the ring. \\
As before, the main difference
between the exact spectra and our mean field results is that in our case,
the angular momentum changes continuously as a function of flux, thus
smearing out the cusps at transitions between different $L$-states. 
This effect, which is due to the approximations made in our calculation,
is similar to the effect an impurity has in the exact spectrum: An impurity
potential, breaking the rotational symmetry, typically smears out the
sharp cusps and opens up gaps at level crossings between different $L$-states.
In the next section, we shall study systematically systems with
such an explicitly symmetry-breaking potential of variable strength.
\PostScript{6}{0.5}{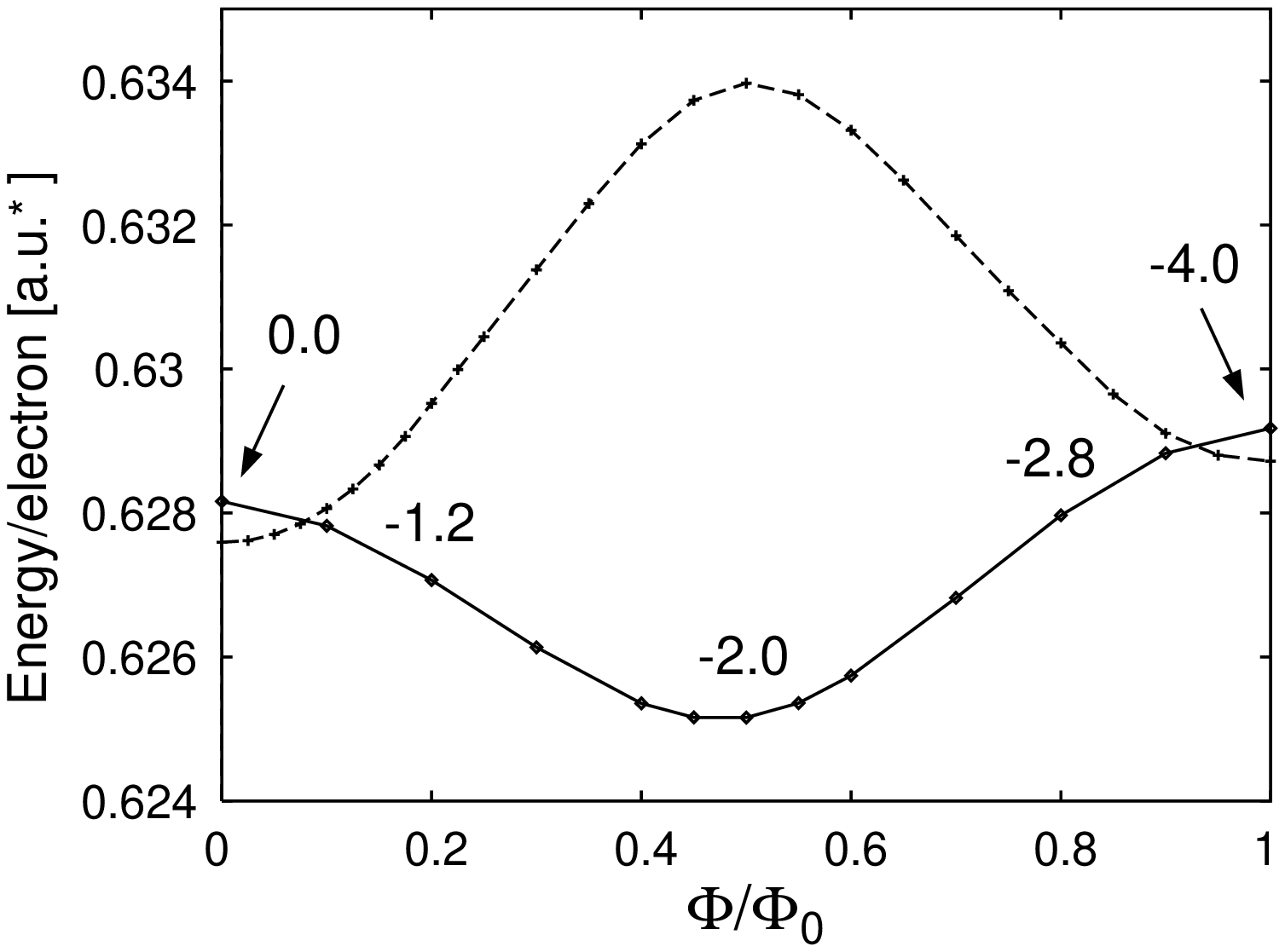}{0.5}{14}{
Four-particle spectrum: Energy per particle of the $S=0$
(solid line) and $S=1$ (dashed line) states of a four-electron
ring with $r_{s,1D}=2.5$, $C_F=10$. The numbers along the $S=0$
curve indicate the total angular momentum $L$ for some values
of $\phi/\phi_0$.}{f2}

As an additional test of our method, we have compared, for given flux,
the numerically computed current (integrated over a cross-section of the ring)
to the derivative of the total energy with respect to flux, cfr. 
Eq.\pref{pc}.
We find very good agreement
(to three decimal places, taking, for example $N=4$, $r_{s, 1D}=2.5$,
$C_F=10$, $S=-1$ and $\phi=0.25 \phi_0$) 
with the theoretical prediction \pref{pc}. 
We thus conclude that CSDFT correctly captures the effect of Aharonov-
Bohm phases picked up by the electrons in the ring.

For completeness, we also include the spectrum of a six-particle ring
with $r_{s, 1D}=2$, $C_F = 7$ (Fig.\ref{f3}), which will be 
discussed in the next section
in the presence of an impurity potential.
\PostScript{6}{0}{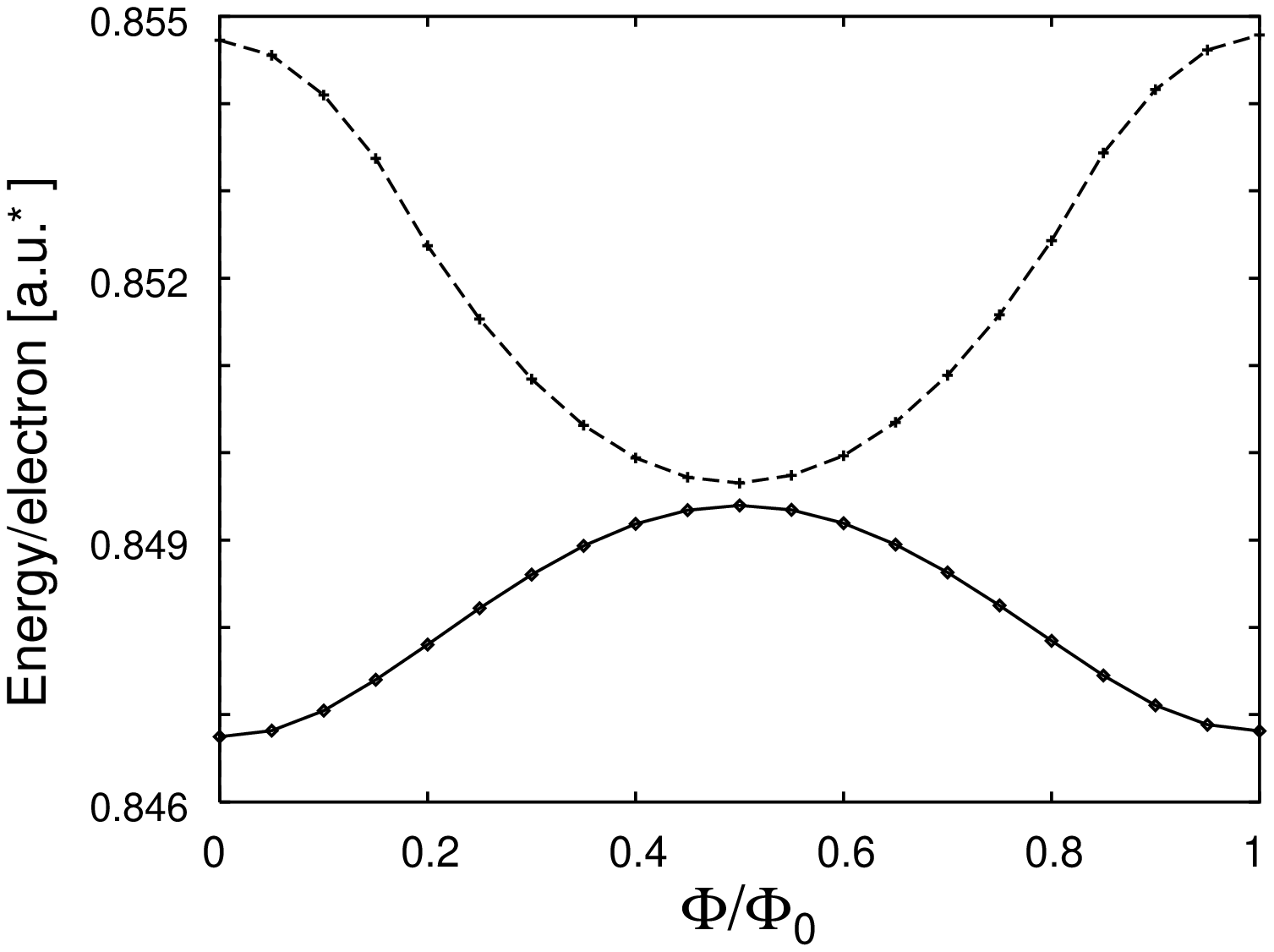}{0.5}{14}{
Six-particle spectrum: Energy per particle of the $S=0$
(solid line) and $S=1$ (dashed line) states of a six-electron
ring with $r_{s,1D}=2$, $C_F=7$.}{f3}
\section{Effects of impurity}
\label{s4}
In this section we will study the effects of a single Gaussian impurity,
given by \pref{imppot},
on the persistent current and charge density of a six-electron ring. 
The main
feature is that the impurity tends to localize the electrons in the ring,
thus reducing the persistent current. This effect has previously been studied
by Cheung \etal\cite{cheung1} in the case of an ideal, one dimensional 
metal ring. 
We will see that our method produces results which are in good qualitative
agreement with those in \cite{cheung1}. \\
There is another mechanism which tends to deform the electron
density: As we shall see, making the ring
more and more narrow, \ie increasing the parameter $C_F$,
creates a charge-density-wave (CDW) in the ring. 
In the following, we shall study
the dependence of the density and persistent current on impurity
strength and on $C_F$.

We start with a direct study of the charge density in a ring containing 
six electrons. Fig.\ref{f4} shows the spin down- and total densities
of a ring with $r_{s, 1D}=2$, $C_F = 7$ and zero flux at various impurity
strengths $V_0$ (with $a=b=2$, see Eq. (\ref{imppot})). 
\PostScript{8}{0}{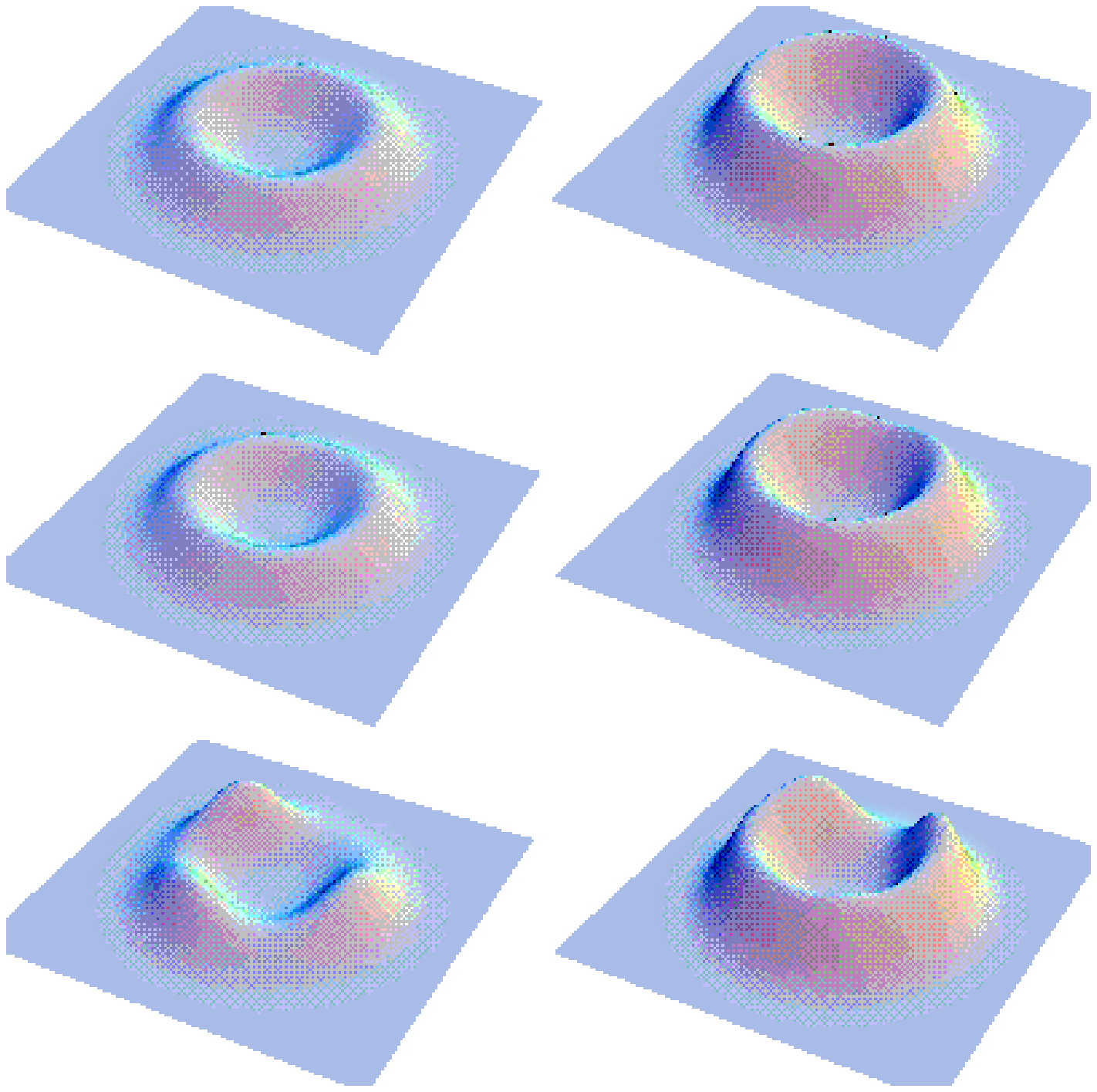}{0.5}{14}{Spin down- (left column) and
total (right column) densities of a six electron ring  with 
$r_{s, 1D}=2$, $C_F = 7$, $\phi=0$, $S=0$ and impurity strengths
(from the top) $V_0=0.0,~0.1,~0.5$. Increasing $V_0$ gradually
localizes the electrons and reduces the density at the location
of the impurity.}{f4}
The spin, $S$, is zero in the
ground state, see Fig.\ref{f3}. 
In this case, the density of both spin components
is homogeneous and rotationally invariant for $V_0=0$ and $0.1$, whereas
one can see that the electrons start to get localized at larger $V_0$.
This localization is accompanied by the formation of a spin-density wave
(SDW), with alternating spin up- and spin down electrons. We also see
that the density at the impurity center decreases with increasing $V_0$.
A more systematic analysis of this effect is presented in Fig.\ref{f5}, 
where we show the total electron density at ($x=R_0$, $y=0$)
as function of impurity strength for
the same ring at three different flux values, and also a more narrow ring
with $C_F = 16$, all normalized by the density at $V_0=0$. The curves
roughly fall on top of each other, independent of $\phi$ and $C_F$ and
seem to fall off exponentially as function of the impurity strength. 
\PostScript{6}{0}{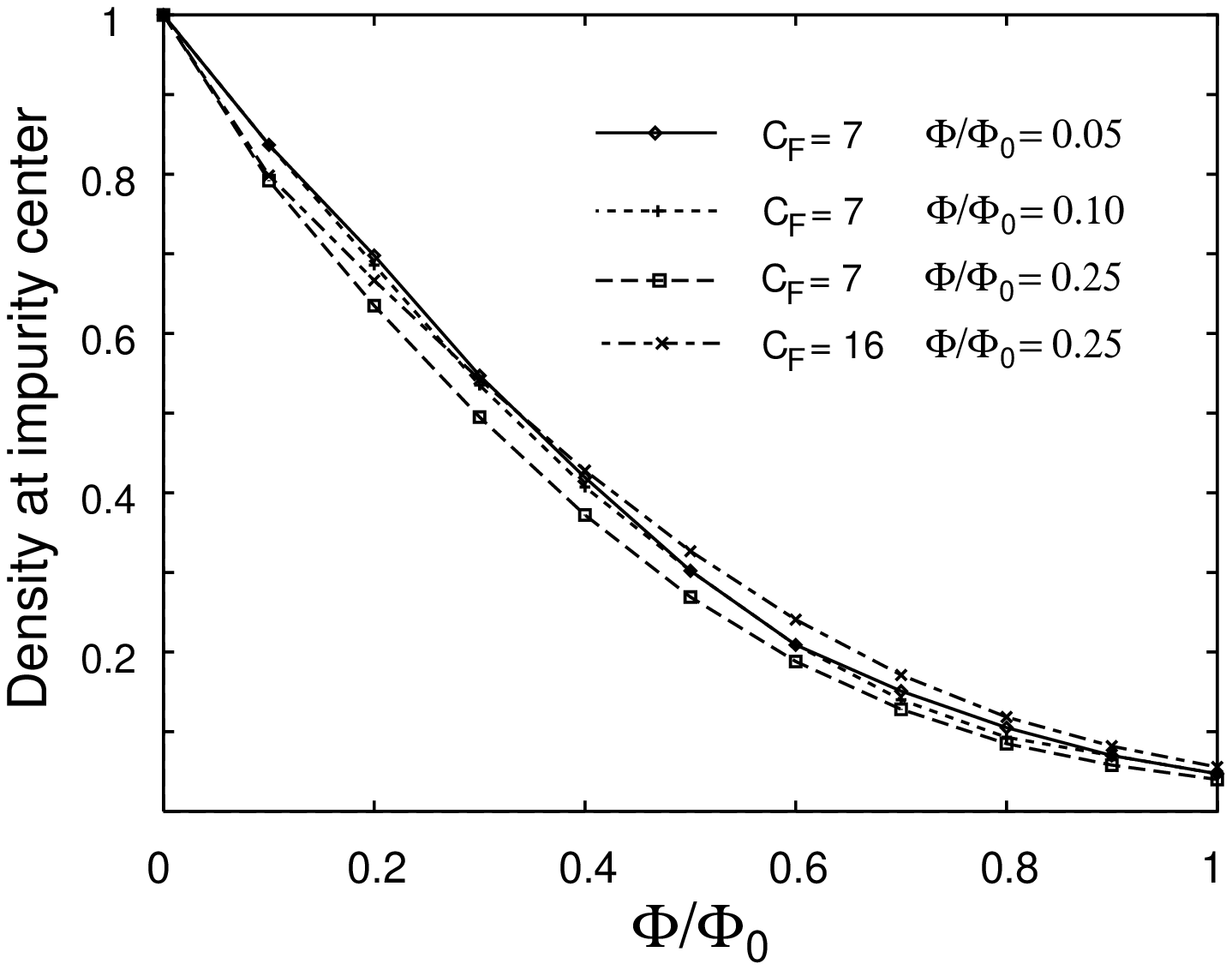}{0.5}{14}{
Density at impurity center vs. impurity strength for a set of
six-electron rings with $r_{s,1D}=2$, $C_F=7$ and $\phi=0.05\phi_0$, 
$\phi=0.1\phi_0$ and $\phi=0.25\phi_0$,
and also $C_F=16$, $\phi=0.25\phi_0$, all in the
$S=0$ state and normalized by their densities at $V_0=0$.
The shape of these curves is roughly independent of flux strength
and of $C_F$.}{f5}

\bigskip
Fig.\ref{f6} shows an example of how the persistent current falls
off with increasing impurity strength $V_0$ ($a=b=2$) in a six-electron
ring ($r_{s,1D}=2$, $C_F=7$) in the $S=1$ state at $\phi=0.25\phi_0$.
This choice of parameters, though not corresponding to the ground state
(see Fig.\ref{f3}), is particularly convenient due to the large
absolute value of the current, making numerical error less significant.
We have checked that other (ground state) sets of parameters give
qualitatively the same behavior.
We note that the form of this curve, with a plateau at small $V_0$,
followed by steeper fall-off, 
is very similar to the result
by Cheung \etal \cite{cheung1}, obtained by numerical diagonalization
of an ideal one-dimensional ring with 20 electrons 
in the tight binding approximation.
\PostScript{6}{0}{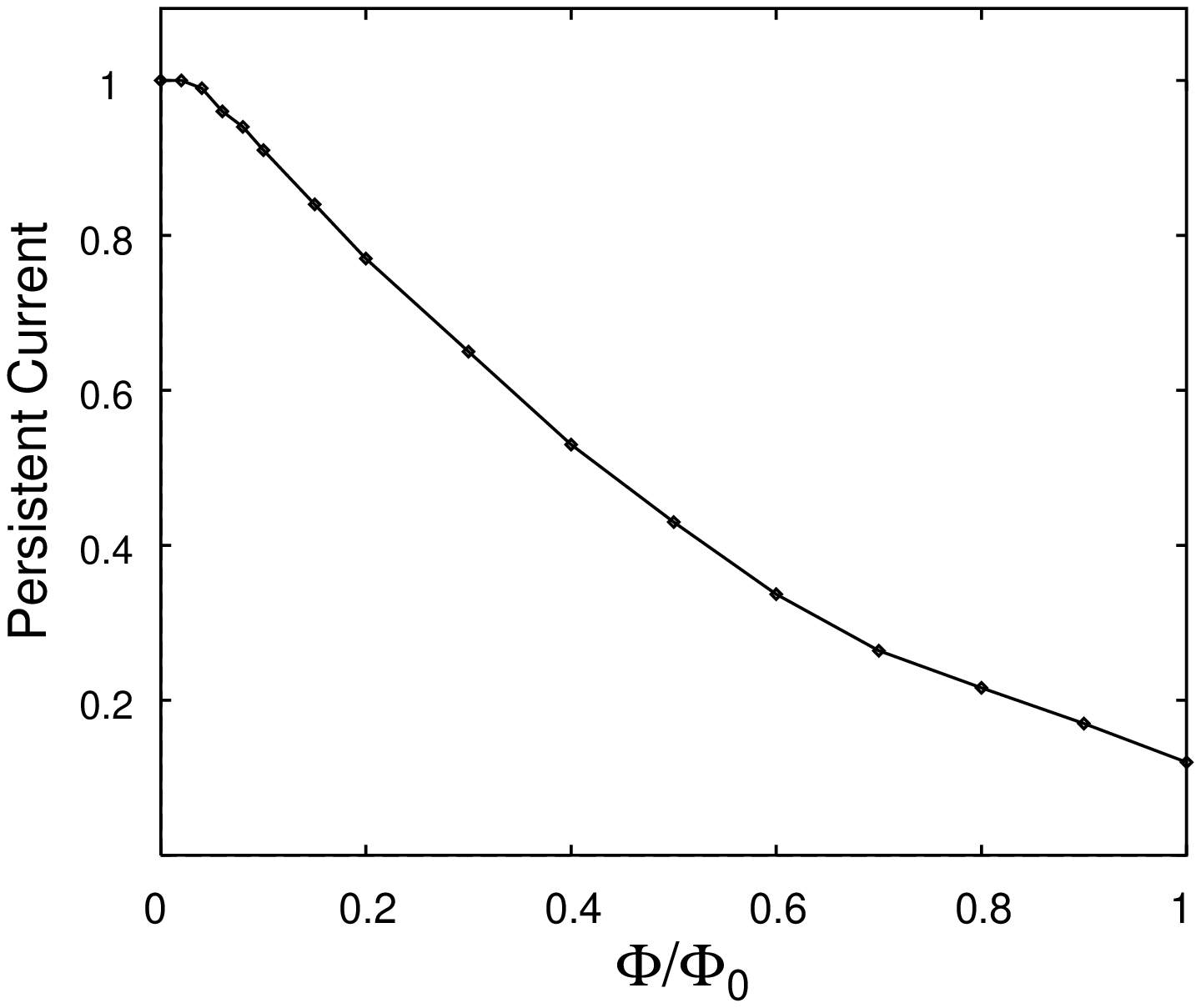}{0}{14}{
Persistent current vs. impurity strength for a six-electron
ring in the $S=1$ state with $r_{s,1D}=2$, $C_F=7$ and 
$\phi=0.25\phi_0$.}{f6}

\bigskip

Finally, we examine another effect which tends to localize the electrons 
in the ring: It turns out that making the ring more one-dimensional,
\ie increasing $C_F$ keeping everything else fixed, gradually localizes
the electrons, creating a strong SDW along with a spatial modulation
of the total charge density (CDW). This is illustrated
in Fig.\ref{f7} where we show density profiles of a six-electron ring with
$r_{s,1D}=2$ at zero flux and without an impurity for three different values
of $C_F$. 
\PostScript{8}{0}{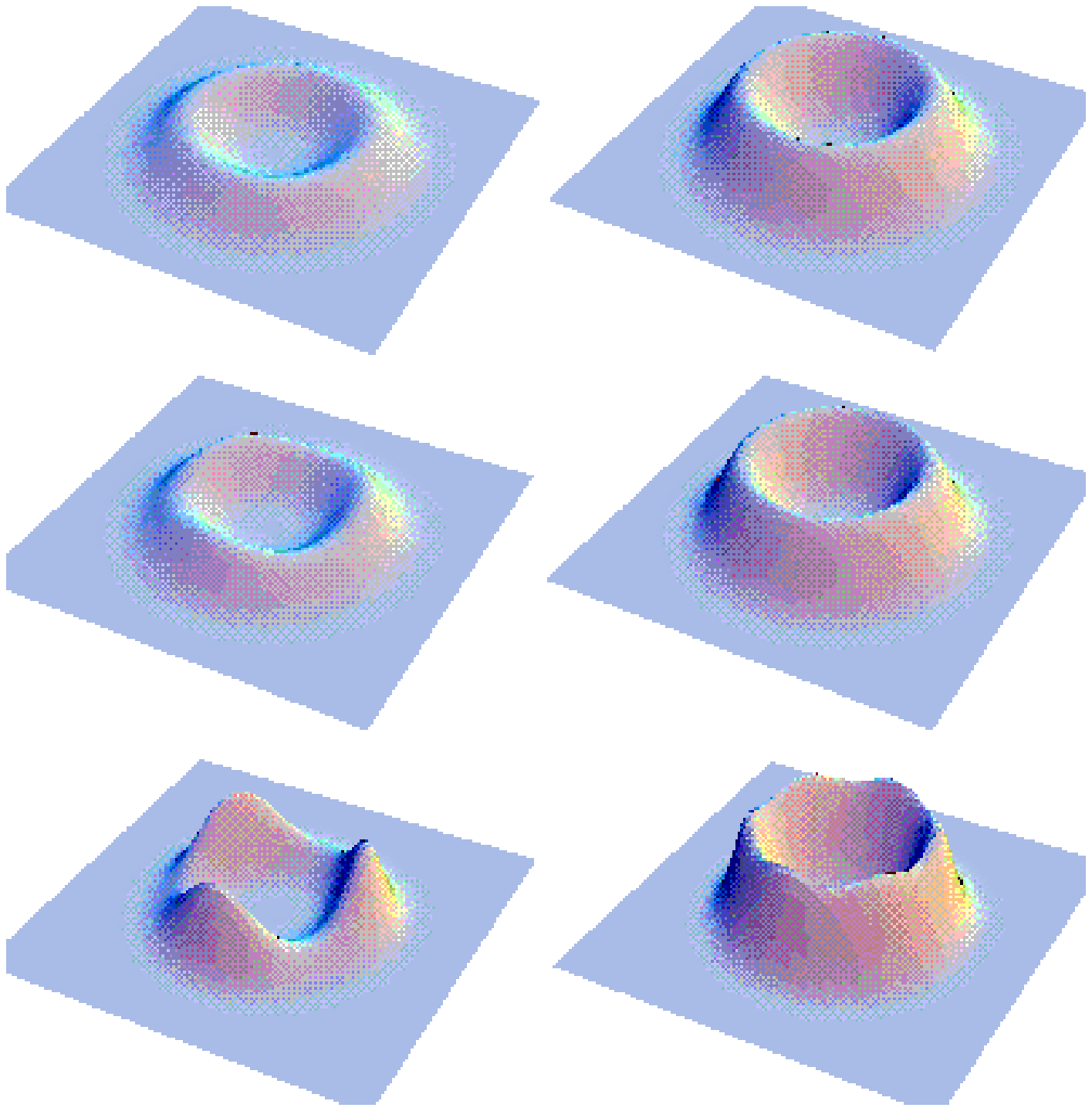}{0.5}{14}{Spin down- (left column) and
total (right column) densities of an impurity-free six electron ring 
with $r_{s, 1D}=2$, $\phi=S=0$ and 
(from the top) $C_F=7,~8,~12$. Increasing $C_F$, \ie narrowing the
confining potential, gradually localizes the electrons.}{f7}
Such localization and antiferromagnetic ordering of the electron spin
in a quasi one-dimensional ring confinement was recently confirmed 
by exact diagonalization studies: the many-body spectra of quantum rings 
with up to 6 electrons could be described by a spin model 
combined with a rigid center-of mass rotation \cite{koskinen2}.
The stronger the ring confinement (\ie the more narrow the 
quasi one-dimensional ring), or the lower the average particle density,
the more pronounced are the effects of localization and formation
of charge density waves in the internal structure of the many-body 
wave function. 

One might expect that an impurity potential would ``pin'' a charge
density wave in the ring, \ie the persistent current of a ring with
localized electrons should be reduced as compared to the non-localized
case. 
We have examined the possibility of such a ``pinning-depinning
transition'' by computing the persistent current at fixed impurity
strength as a function of $C_F$. Fig.\ref{f8} shows the result for several
different values of the impurity strength $V_0$. We see that the current
indeed decreases as the ring becomes more narrow; however, there is no
``abrupt'' transition since, as we have seen, localization happens
gradually with increasing $C_F$. 
\PostScript{6}{0}{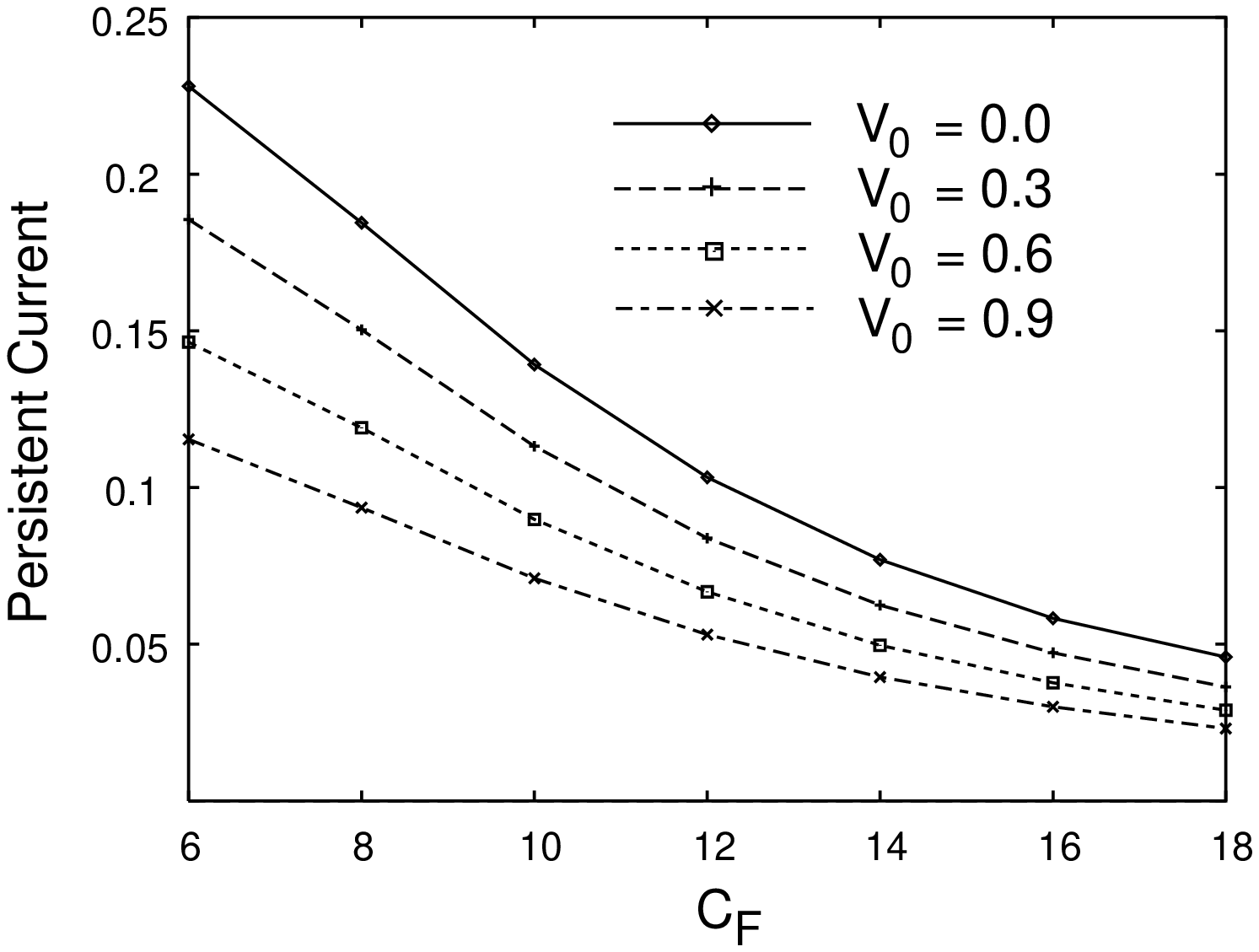}{0}{14}{
Persistent current vs. $C_F$ for a six-electron ring with
$S=0,~r_{s,1D}=2,~\phi=0.25 \phi_0$
at various impurity strengths,
$V_0=0.0,~ 0.3,~ 0.6,~ 0.9$ (from the top) and $a=2.5$, $b=0.5$.}{f8}
Note the interesting scaling behavior suggested by Fig.\ref{f8}: 
The ratio
between any two of the curves is just a constant, independent of $C_F$;
in particular, the ratio $I(V_0)/I(V_0=0)$ is independent of $C_F$
for any $V_0$.
Also note that the persistent current decreases with increasing $C_F$
even in the {\em zero} impurity case, thus making no qualitative
distinction between a ``clean'' and a ``dirty'' ring. 
This may again be due to the explicit
symmetry breaking by the LSDA which, as we have discussed previously,
in a sense mimics disorder even in the case $V_0=0$.

\section{Conclusion}
\label{s5}
The numerical analysis presented here demonstrates that
the current-spin-density-functional formalism provides
a suitable tool for describing the effects of Aharonov-Bohm
phases and impurities in realistic quantum rings. In particular, we
have shown that it reproduces the main qualitative features
of the many-body spectra and persistent currents, taking
into account the effects of interactions, spin and deviations
from perfect one-dimensionality. 
Furthermore, we have shown that the persistent current
in the ring may be suppressed by narrowing the confining
potential at fixed impurity strength.

The main difference between
this model and exact diagonalization results is that the LSDA
introduces a breaking of the rotational symmetry in the ring
even in the impurity-free case. One may hope that this is
not a problem when describing 
experimentally realizable quantum rings, as 
a certain amount of disorder and non-perfect symmetry 
is expected to be present in any realistic system.

With the present experimental progress in fabricating
and studying few-electron quantum rings, the methods
described here may turn out useful for suggesting and
describing future experiments.

\vskip 3mm
\noi
{\bf ACKNOWLEDGEMENT}:  
This work was financially supported by the Academy of Finland,
the TMR programme of the European Community under contract 
ERBFMBICT972405, the ``Bayerische Staatsministerium f\"ur 
Wissenschaft, Forschung und Kunst'', and the NORDITA Nordic
project ``Confined electronic systems''.

\end{multicols}

\end{document}